\documentclass[pamm,a4paper,fleqn,oneside]{w-art}
\usepackage{times}
\usepackage{w-thm}
\usepackage{hyperref}
\theoremstyle{plain}

\theoremstyle{definition}

\usepackage[]{graphicx}
\chardef\bslash=`\\ 

\begin{document}

\begin{minipage}[t]{180mm}
\thispagestyle{empty}
\vspace{20mm}

\begin{center}
{\Large\bf Patterns in Wigner-Weyl approach}

\vspace{20mm}

{\large\bf Antonina N. Fedorova, Michael G. Zeitlin}

\vspace{20mm}

Mathematical Methods in Mechanics Group \\

Institute of Problems of Mechanical Engineering (IPME RAS)\\ 

Russian Academy of Sciences \\

Russia, 199178, St. Petersburg, V.O., Bolshoj pr., 61\\

zeitlin@math.ipme.ru, anton@math.ipme.ru\\
         
http://www.ipme.ru/zeitlin.html\\

http://mp.ipme.ru/zeitlin.html

\vspace{20mm}
{\bf Abstract}

\vspace{10mm}
\begin{tabular}{p{100mm}}

We present a family of methods, which can describe behaviour of 
quantum ensembles
and demonstrate the creation of nontrivial (meta) stable
states (patterns), localized, chaotic, entangled or decoherent 
from basic localized modes
in collective models arising from the 
quantum hierarchy 
of Wigner-von Neumann-Moyal equations.

\vspace{10mm}

Presented: GAMM Meeting, 2006, Berlin, Germany.
\vspace{5mm}

Published: Proc. Appl. Math. Mech. (PAMM), {\bf 6}, 627-628, Wiley-VCH, 2006.\\

\vspace{5mm}

 {\bf DOI} 10.1002/pamm.200610294

\end{tabular}

\end{center}
\end{minipage}
\newpage

\title{Patterns in Wigner-Weyl approach}


\author{Antonina N. Fedorova}
\author{Michael G. Zeitlin\footnote{Corresponding
     author: e-mail: {\sf zeitlin@math.ipme.ru}, http://www.ipme.ru/zeitlin.html,
  http://mp.ipme.ru/zeitlin.html}}
\address[]{IPME RAS, St.~Petersburg, V.O. Bolshoj pr., 61, 199178, Russia}

\begin{abstract}
We present a family of methods, which can describe behaviour of 
quantum ensembles
and demonstrate the creation of nontrivial (meta) stable
states (patterns), localized, chaotic, entangled or decoherent 
from basic localized modes
in collective models arising from the 
quantum hierarchy 
of Wigner-von Neumann-Moyal equations.
\end{abstract}
\maketitle                   

It is widely known that the currently available experimental techniques in the area of 
quantum physics as a whole and in that of quantum computations in particular, as well 
as the present level of understanding of phenomenological models, outstripped
the actual level of mathematical/theoretical description.
Considering, for example, the problem of describing the realizable states
one should not expect that planar waves and (squeezed) gaussian coherent states 
would be enough to characterize such complex systems as 
qCPU (quantum Central Processor Unit)-like devices.
Complexity of the set of relevant states, including entangled (chaotic) 
states is still far from being clearly understood and moreover from being realizable.
Our main motivations arise from the following 
general questions:
How can we represent well localized and reasonable state in mathematically correct form?
Is it possible to create entangled and other relevant states by means of these new building blocks?
A starting point for us is a possible model for (continuous) ``qudit'' with subsequent 
description of the whole zoo of possible realizable (controllable) states/patterns 
which may be useful from the point of view of quantum 
experimentalists and engineers.
Effects we are interested in are:
(a)hierarchy of internal/hidden scales (time, space, phase space);
(b)non-perturbative multiscales: from slow to fast contributions,
from the coarser to the finer level of resolution/decomposition;
(c)coexistence of hierarchy of multiscale dynamics with transitions between scales.
(d)realization of the key features of the complex quantum 
world such as the existence of chaotic and/or entangled 
states with possible destruction in ``open/dissipative'' regimes due to interactions with
quantum/classical environment and transition to decoherent states.
By localized states (localized modes) 
we mean the building blocks for solutions or generating modes which 
are localized in maximally small region of the phase 
(as in c-(classical) as in q(quantum)-case) space.
By an entangled/chaotic pattern we mean some solution (or asymptotics of solution) 
which has random-like distributed energy (or information) spectrum in a full domain of definition. 
In quantum case we need to consider additional entangled-like patterns, roughly speaking,
which cannot be separated into pieces of sub-systems.
By a localized pattern (waveleton) 
we mean (asymptotically) (meta) stable solution localized in a 
relatively small region of the whole phase space (or a domain of definition). 
In this case the energy is distributed during some time (sufficiently large) 
between only a few  localized modes (from point 1). 
We believe it to be a good model for plasma in a fusion state (energy confinement)
or a model for quantum continuous ``qubit'' or a result of the decoherence process
in open quantum system when the full entangled state degenerates 
into localized (quasiclassical) pattern.
In this paper we consider the calculations of the Wigner functions
$W(p,q,t)$ (WF) corresponding
to the classical polynomial Hamiltonian $H(p,q,t)$ as the solution
of the Wigner-von Neumann equation:
$\quad
i\hbar W_t = H * W - W * H
$
and related Wigner-like equations for different ensembles.
According to the Weyl transform, a quantum state (wave function or density 
operator $\rho$) corresponds
to the Wigner function, which is the analogue in some 
sense of classical phase-space distribution [1].
Wigner equation is a result of the Weyl transform 
or ``wignerization'' of von Neumann equation for density matrix.
The difference between classical and quantum case is concentrated
in the structure of the set of operators included in the set-up and, surely,
depends on the method of quantization.
But, in the naive Wigner-Weyl approach for quantum case the symbols of operators  
play the same role as usual functions in classical case. 
To move from $c$- to $q$-case, let us start from the second quantized 
representation for an algebra of observables 
$A=(A_0,A_1,\dots,A_s,...)$
in the standard form
$
A=A_0+\int dx_1\Psi^+(x_1)A_1\Psi(x_1)+\dots+
(s!)^{-1}\int dx_1\dots dx_s\Psi^+(x_1)\dots
\Psi^+(x_s)A_s\Psi(x_s)\dots\Psi(x_1)+\dots
$
N-particle Wigner function
$
W_s(x_1,\dots,x_s)=\int dk_1\dots dk_s{\rm exp}\big(-i\sum^s_{i=1}k_ip_i\big)
{\rm Tr}\rho\Psi^+\big(q_1-\frac{1}{2}\hbar k_1\big)\dots
$,
$
\Psi^+\big(q_s-\frac{1}{2}\hbar k_s\big)\Psi\big(q_s+\frac{1}{2}\hbar 
k_s\big)\dots
\Psi\big(q_1+\frac{1}{2}\hbar k_s\big)
$
allow us to consider them as some quasiprobabilities and provide useful bridge
between c- and q-cases:
$
<A>={\rm Tr}\rho A=\sum^{\infty}_{s=0}(s!)^{-1}\int\prod_{i=1}^s 
d\mu_iA_s(x_1,\dots,x_s)
W_s(x_1,\dots,x_s).
$
The full description for quantum ensemble can be done by the whole hierarchy
of functions (symbols):
$
W=\{W_s(x_1,\dots,x_s), s=0,1,2\dots\}
$
So, we may consider the following q-hierarchy as the result of ``wignerization'' 
procedure for c-BBGKY one,
where the partial Wigner functions
are solutions of proper Wigner equations:

\begin{eqnarray}
\frac{\partial W_n}{\partial t}=-\frac{p}{m}\frac{\partial W_n}{\partial q}+
\sum^{\infty}_{\ell=0}\frac{(-1)^\ell(\hbar/2)^{2\ell}}{(2\ell+1)!}
\frac{\partial^{2\ell+1}U_n(q)}{\partial q^{2\ell+1}}
\frac{\partial^{2\ell+1}W_n}{\partial p^{2\ell+1}}.
\end{eqnarray}
We obtain our multiscale/multiresolution representations for solutions of Wig\-ner-like equations
via a variational-wavelet approach. 
We represent the solutions as 
decomposition into localized eigenmodes (regarding action of affine group, i.e.
hidden symmetry of the underlying functional space of states) 
related to the hidden underlying set of scales: 
$
W_n(t,q,p)=\displaystyle\bigoplus^\infty_{i=i_c}W^i_n(t,q,p),
$
where value $i_c$ corresponds to the coarsest level of resolution
$c$ or to the internal scale with the number $c$ in 
the full multiresolution decomposition (MRA)
of the underlying functional space ($L^2$, e.g.):
$
V_c\subset V_{c+1}\subset V_{c+2}\subset\dots
$
We introduce the Fock-like space structure (in addition to the standard one, 
if we consider second-quantized case) on the whole space of internal hidden scales
$
H=\bigoplus_i\bigotimes_n H^n_i
$
for the set of n-partial Wigner functions (states):
$
W^i=\{W^i_0,W^i_1(x_1;t),\dots,
W^i_N(x_1,\dots,x_N;t),\dots\},
$
where
$W_p(x_1,\dots, x_p;t)\in H^p$,
$H^0=C,\quad H^p=L^2(R^{6p})$ (or any different proper functional spa\-ce), 
with the natural Fock space like norm: 
$
(W,W)=W^2_0+
\sum_{i}\int W^2_i(x_1,\dots,x_i;t)\prod^i_{\ell=1}\mu_\ell.
$
We consider $W=W(t)$,
$W\in L^2(R)$, via
multiresolution decomposition which naturally and efficiently introduces 
the infinite sequence of the underlying hidden scales [2].
We have the contribution to
the final result from each scale of resolution from the whole
infinite scale of spaces.
The closed subspace
$D_j ($j$\in {\bf Z})$ corresponds to  the level $j$ of resolution, 
or to the scale $j$.
Then we have the following decomposition:
$
\{W(t)\}=\bigoplus_{-\infty<j<\infty} D_j 
=\overline{V_c\displaystyle\bigoplus^\infty_{j=0} D_j},
$
in case when $V_c$ is the coarsest scale of resolution.
As a result the solution of this hierarchy
has the 
following mul\-ti\-sca\-le or mul\-ti\-re\-so\-lu\-ti\-on decomposition via 
nonlinear high\--lo\-ca\-li\-zed eigenmodes 
\begin{eqnarray}
&&W(t,x_1,x_2,\dots)=
\sum_{(i,j)\in Z^2}a_{ij}U^i\otimes V^j(t,x_1,\dots),\\
&&V^j(t)=
V_N^{j,slow}(t)+\sum_{l\geq N}V^j_l(\omega_lt), \ \omega_l\sim 2^l, 
\quad U^i(x_s)=
U_M^{i,slow}(x_s)+\sum_{m\geq M}U^i_m(k^{s}_mx_s), \ k^{s}_m\sim 2^m,\nonumber
\end{eqnarray}
which corresponds to the full multiresolution expansion in all underlying time/space 
scales.
It should be noted that such representations 
give the best possible localization
properties in the corresponding (phase)space/time coordinates. 
Numerical calculations are based on compactly supported
wavelets and wavelet packets and on evaluation of the 
accuracy on 
the level $N$ of the corresponding cut-off of the full system 
regarding Fock-like norm above:
$
\|W^{N+1}-W^{N}\|\leq\varepsilon.
$



\begin{vchfigure}
\begin{center}
\begin{tabular}{cc}
\includegraphics*[width=65mm]{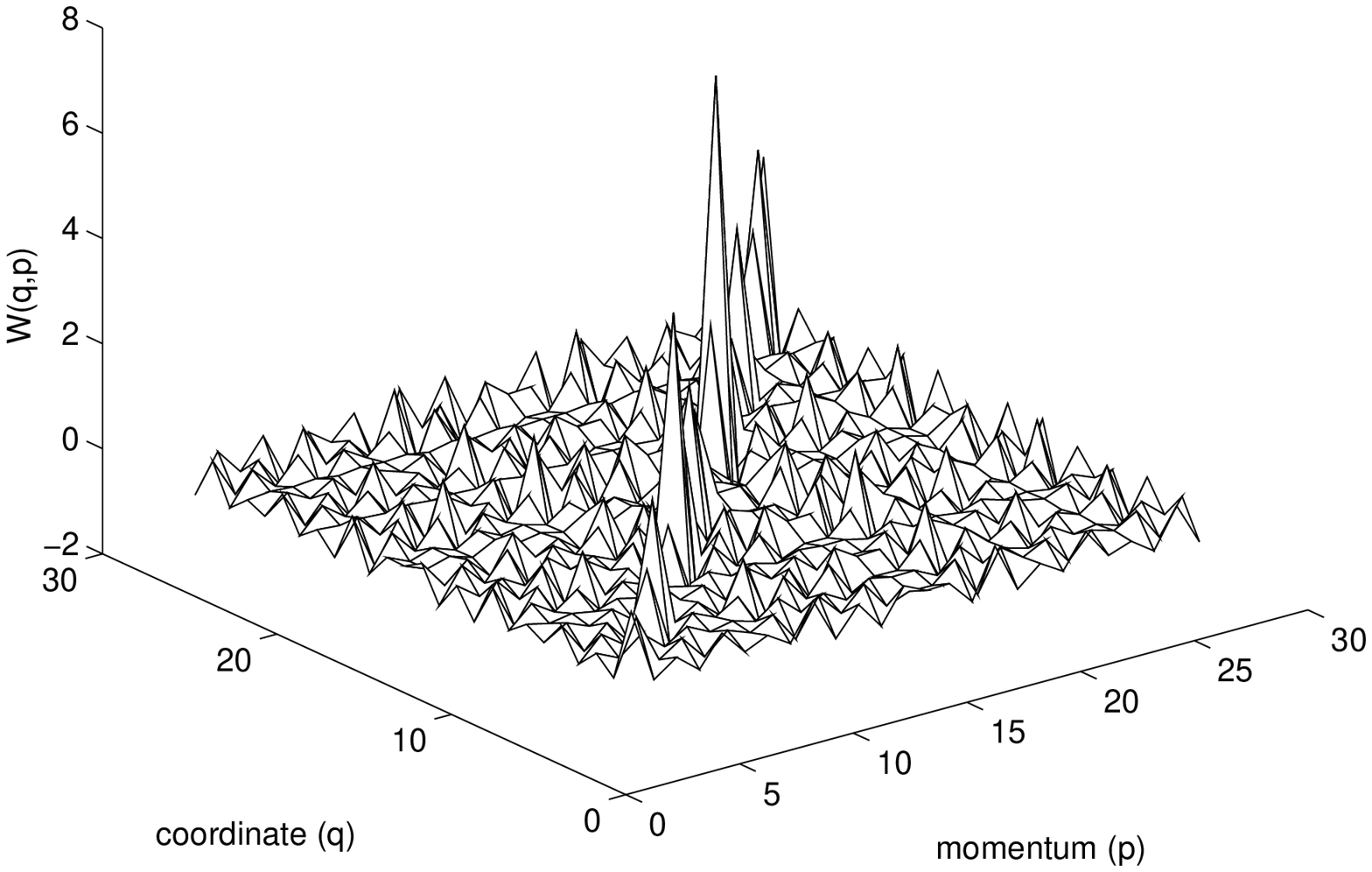} &
\includegraphics*[width=65mm]{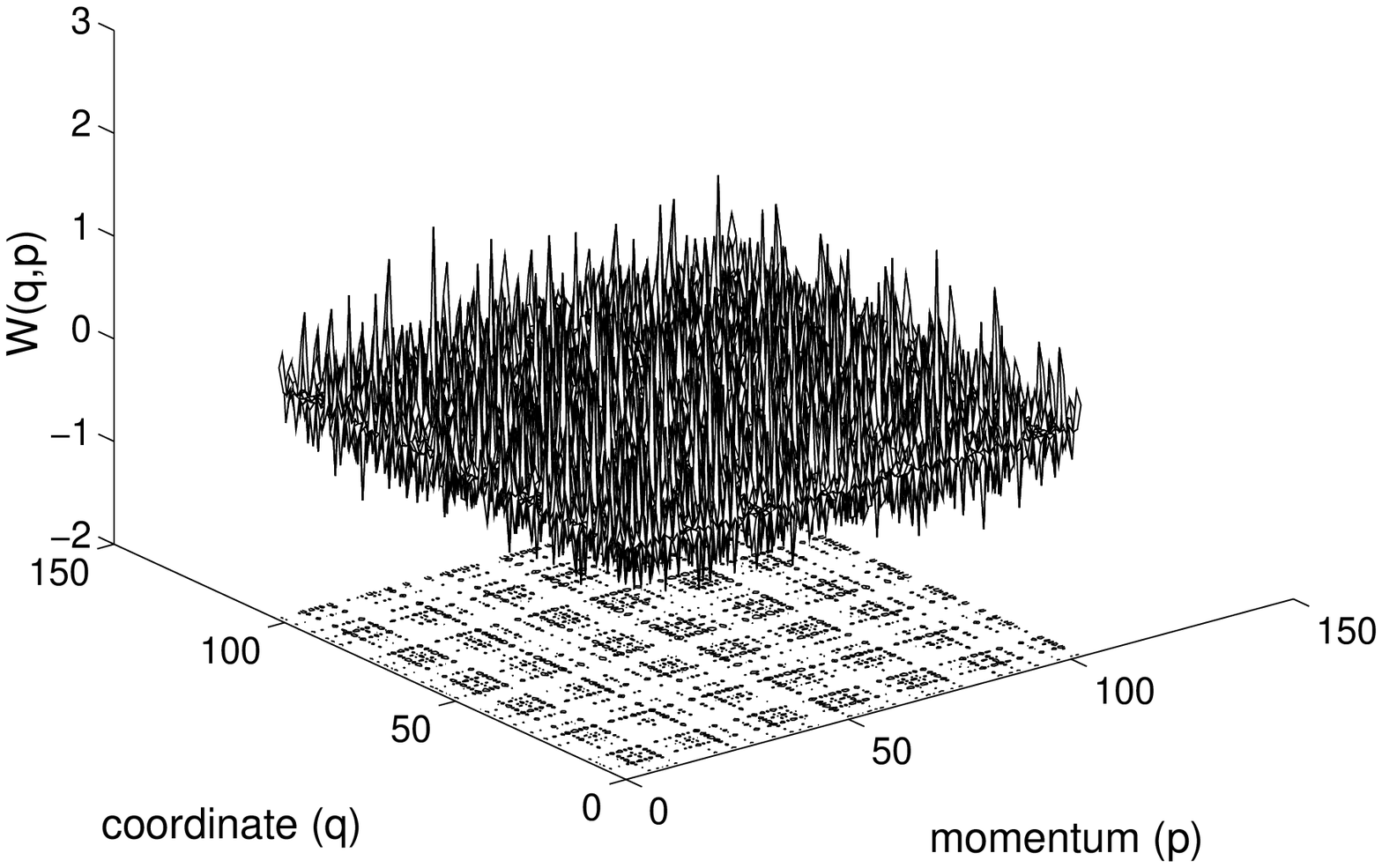}
\end{tabular}
\end{center}
\vchcaption{Localized pattern or waveleton Wigner function (left);
entangled-like Wigner function (right).}
\end{vchfigure}

The numerical simulation demonstrates the formation of different (stable) pattern or orbits 
generated by internal hidden symmetry from
high-localized structures.
Our (nonlinear) eigenmodes are more realistic for the modeling of 
nonlinear classical/quantum dynamical process  than the corresponding linear gaussian-like
coherent states. Here we mention only the best convergence properties of the expansions 
based on wavelet packets, which  realize the minimal Shannon entropy property
and the exponential control of convergence of expansions like (2) 
based on the norm above.
Fig.~1 (left) corresponds to (possible) result of superselection
(einselection) [1] after decoherence process started from entangled state (Fig.~1, right).

\end{document}